\documentclass[twocolumn,showpacs,preprintnumbers,amsmath,amssymb]{revtex4}

\usepackage{graphicx}
\usepackage{bm}

\usepackage{color}

\begin{document}

\title{Dense granular flows: \\interpolating between grain inertia and fluid viscosity based constitutive laws}

\author{Pierre Rognon}
\author{Cyprien Gay}

\email{cyprien.gay@univ-paris-diderot.fr}

\affiliation{%
Centre de Recherche Paul Pascal, CNRS UPR~8641 - Av. Dr. Schweitzer, Pessac, France\\
Mati\`{e}re et Syst\`{e}mes Complexes, Universit\'{e} Paris-Diderot - Paris 7, CNRS UMR~7057 - Paris, France
}%

\date{\today}

\begin{abstract}
A scalar constitutive law was recently obtained for dense granular flows
from a two-grain argument,
both in the inertial regime (grain inertia)
and in the viscous regime.
As the resulting law is not exactly the same in both regimes,
we here provide an expression for the crossover between both regimes.
\end{abstract}

\maketitle

\newcommand{\hide}[1]{#1}

\newcommand{\hs}{\hspace{0.8cm}}
\newcommand{\dd}{{\rm d}}
\newcommand{\be}{\begin{equation}}
\newcommand{\ee}{\end{equation}}
\newcommand{\bee}{\begin{eqnarray}}
\newcommand{\eee}{\end{eqnarray}}

\newcommand{\transp}[1]{{#1}^T}
\newcommand{\trace}{{\rm tr}}

\newcommand{\gd}{\dot{\gamma}}
\newcommand{\philiq}{\phi}
\newcommand{\phisol}{\psi}
\newcommand{\phisolzero}{\phisol_0}
\newcommand{\visc}{\eta}
\newcommand{\viscapp}{\visc_{\rm app}}
\newcommand{\rayon}{d}
\newcommand{\gapmax}{\Delta}
\newcommand{\masse}{m}
\newcommand{\gap}{h}
\newcommand{\gapmin}{h_{\rm min}}

\newcommand{\contactdensity}{c}
\newcommand{\contactenergy}{{\cal E}_{\rm c}}
\newcommand{\slidingenergy}{{\cal E}_{\rm sl}}
\newcommand{\slidingdistance}{d_{\rm sl}}
\newcommand{\fn}{F_N}
\newcommand{\fnapp}{F_N^{\rm app}}
\newcommand{\fnsep}{F_N^{\rm sep}}
\newcommand{\ft}{F_T}
\newcommand{\vn}{v_N}
\newcommand{\vt}{v_T}
\newcommand{\hmin}{h_{min}}
\newcommand{\coordinance}{z}
\newcommand{\partofdeformation}{f}
\newcommand{\sectioneffective}{A} 
\newcommand{\period}{T}
\newcommand{\periodVIS}{\period^{\rm v}}
\newcommand{\periodINE}{\period^{\rm i}}
\newcommand{\ratio}{\cal {R}}

\newcommand{\Tmicro}{T_{micro}}
\newcommand{\Tsl}{T_{sl}}
\newcommand{\Tapp}{T^{app}}
\newcommand{\Tsep}{T^{sep}}
\newcommand{\Tcar}{T_{car}}
\newcommand{\Tc}{T^\star}
\newcommand{\Tstokes}{T_{Stokes}}
\newcommand{\Tfv}{T_{\rm fv}}
\newcommand{\Tinertiel}{T_{\rm gi}}
\newcommand{\Igi}{I_{\rm gi}}
\newcommand{\Ifv}{I_{\rm fv}}
\newcommand{\Ifi}{I_{\rm fi}}

\newcommand{\si}{\sigma}
\newcommand{\sic}{\sipr^\star}
\newcommand{\sidev}{\tau}
\newcommand{\sipr}{P}

\section{Introduction}

The deformations of granular materials
are usually categorized into
quasistatic deformations, dense flows 
and collisional flows~\cite{Coussot05,GDR04}.

In the quasistatic and dense regimes,
the shear-rate dependence
can be expressed~\cite{GDR04} 
in terms of a parameter $I=T\gd$,
where $T$ is the typical time for a single grain, 
accelerated by the pressure $P$ (or force $Pd^2$), 
to move over a distance comparable to its own size $d$. 
For dry grains of mass $\masse$, 
the surrouding fluid can often be neglected 
and the pressure is resisted only by grain inertia (``gi'')~\cite{Dacruz05a}:
\be
\label{Eq:T_gi}
\masse \, \frac{\rayon}{\Tinertiel^2} \simeq \sipr\,\rayon^2
\hs{i.e.,}\hs
\Igi=\gd\Tinertiel=\gd\,\sqrt{\frac{\masse}{\sipr\,\rayon}}
\ee
\noindent For granular pastes, if the grain inertia is negligible 
as compared to the viscosity of the surrounding fluid (``fv''), 
the force $Pd^2$ exerced by the pressure 
is resisted by a typical Stokes viscous drag force 
felt by a single grain moving at the velocity $d/\Tfv$~\cite{Cassar05}: 
\be
\label{Eq:T_fv}
 \visc\,\rayon\,\frac{\rayon}{\Tfv} \simeq \sipr\,\rayon^2 
\hs{i.e.,}\hs
\Ifv=\gd\Tfv = \gd\,\frac{\visc}{\sipr}
\ee
Empirically, the dimensionless parameter $I$
has been used in the form of a frictional constitutive law,
in terms of the ratio between 
the shear stress $\sidev$ and the pressure $\sipr$:
\be
\label{Eq:mu_I_frictional}
\frac{\sidev}{\sipr} =\mu(I)
\ee
Empirically again, a universal function $\mu(I)$
seems to account for existing data
in both regimes~\cite{Cassar05}:
\be
\label{Eq:mu_I_cassar}
\frac{\mu(I)}{\mu_c}-1=\frac{\mu_2/\mu_c-1}{I_0/I+1}
\ee

\begin{figure}
\begin{center}
\resizebox{0.8\columnwidth}{!}{%
\input{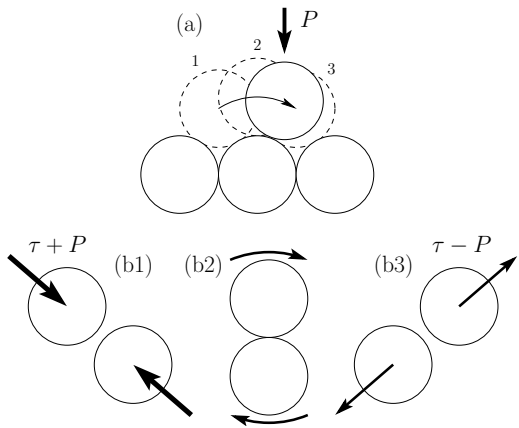}
}
\end{center}
\caption{Schematic evolution within the granular material during shear. 
(a) One grain is transported quasistatically
{from} position 1 to position 2, then falls into position 3
due to the applied pressure $\sipr$.
In a recent work~\cite{Rognon08rheograins},
we rather consider a pair of grains
during the period of time when they are close neighbours (b1-3).
First, the deviatoric (typically, shear) component of the stress, $\sidev$,
helps the pressure $\sipr$ establish the contact (b1).
Then, the contact rotates due to the overall material deformation (b2).
Finally, the deviatoric stress overcomes the pressure
to break the contact (b3).
Because the pressure is compressive in a non-cohesive granular material,
the typical magnitude of the force transmitted between both grains
is stronger when the contact forms than when it breaks.
}
\label{Fig:one_grain_two_grains}
\end{figure}

Recently, a two-grain argument 
(see Fig.~\ref{Fig:one_grain_two_grains}),
based on the same physics,
led to a clear distinction between inter-grain
approach and separation times~\cite{Rognon08rheograins}:
\be
\label{Eq:shear_rate_related_to_contact_lifetime}
\frac{1}{\gd} \simeq \Tapp + \Tsep
\ee

As a result, two different functions arose
for the regime where the grain inertia dominates
and for the regime governed by the fluid viscosity:
\bee
\label{Eq:rheo_gi_complete}
\frac{1}{1+\sqrt{\frac{\sidev-\sipr}{\sidev+\sipr}}}\,
\sqrt{\frac{\sidev}{\sipr}-1}
&=& \Igi \\
%
\label{Eq:rheo_fv}
\frac{\sidev}{\sipr} - \frac{\sipr}{\sidev}
&=& \Ifv
\eee
They do not include the saturation
($\mu(I)\rightarrow\mu_2$ at large $I$)
that is present in Eq.~(\ref{Eq:mu_I_cassar})
and which reflects the onset of the collisional regime.\newline

In the present note, we derive an interpolation
between both regimes represented
by Eqs.~(\ref{Eq:rheo_gi_complete}) and~(\ref{Eq:rheo_fv}).

\begin{figure}
\begin{center}
\resizebox{1.0\columnwidth}{!}{%
\input{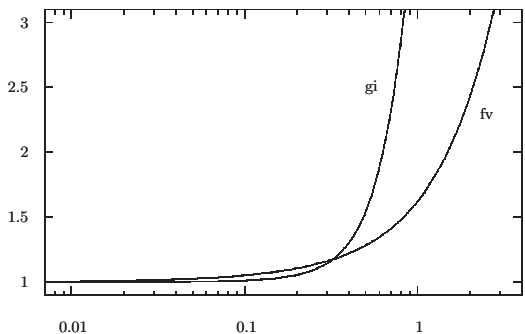}
}
\end{center}
\caption{Raw predictions of the two-particle 
argument~\cite{Rognon08rheograins}
in the regimes dominated by grain inertia and by fluid viscosity,
see Eqs.~(\ref{Eq:rheo_gi_complete}-\ref{Eq:rheo_fv}).
}
\label{Fig:fonctions}
\end{figure}

\section{Crossover from grain inertia to fluid viscosity}

The regimes discussed here are dominated
either by the grain inertia or by the fluid viscosity.
In other or in future experiments,
it may happen that the system lie in the crossover
between regime ``gi'' and regime ``fv''.
In the present section, we discuss
how it is possible to interpolate
between both behaviours.

Cassar {\em et al.}~\cite{Cassar05}
proposed a universal behaviour, given by Eq.~(\ref{Eq:mu_I_cassar}),
which they showed to be compatible both with the data
in the ``gi'' regime and with that in the ``fv'' regime.
Nevertheless, the dimensionless parameter $I$
in Eq.~(\ref{Eq:mu_I_cassar}) does not have the same meaning,
as it is given either by Eq.~(\ref{Eq:T_gi}) 
or by Eq.~(\ref{Eq:T_fv}).
Hence, 
in the crossover region, there is no obvious interpolation between both definitions of $I$.

In the present approach, we not only have this difficulty with $I$,
but we additionally have two different constitutive relations,
namely Eqs.~(\ref{Eq:rheo_gi_complete}) and~(\ref{Eq:rheo_fv}).
We therefore need to go back to the equation
introduced by Courrech du Pont~\cite{Courrech03}
for freely falling grains in avalanches,
and used by Cassar {\em et al.}~\cite{Cassar05}
with a pressure $\sipr$.
Omitting all numerical coefficients:
\be
\masse\,\dot{v}\simeq\sipr\,\rayon^2-\rayon\,\visc\,v
\ee
where $v$ is the grain velocity,
$\sipr\,\rayon^2$ is the typical force
resulting from the pressure,
and $\rayon\,\visc\,v$ is the Stokes drag force
of the grain in the fluid.
With vanishing initial velocity $v(0)$ and position $x(0)$,
we can derive the grain position:
\be
\label{Eq:solution_equadif}
x(t)=\frac{\rayon\,\sipr}{\visc}\,t
-\frac{\masse\,\sipr}{\visc^2}\,
\left(1-\exp\left\{-\frac{\visc\,\rayon}{\masse}\,t\right\}\right)
\ee

\hide{
\begin{figure}
\begin{center}
\resizebox{1.0\columnwidth}{!}{%
\includegraphics{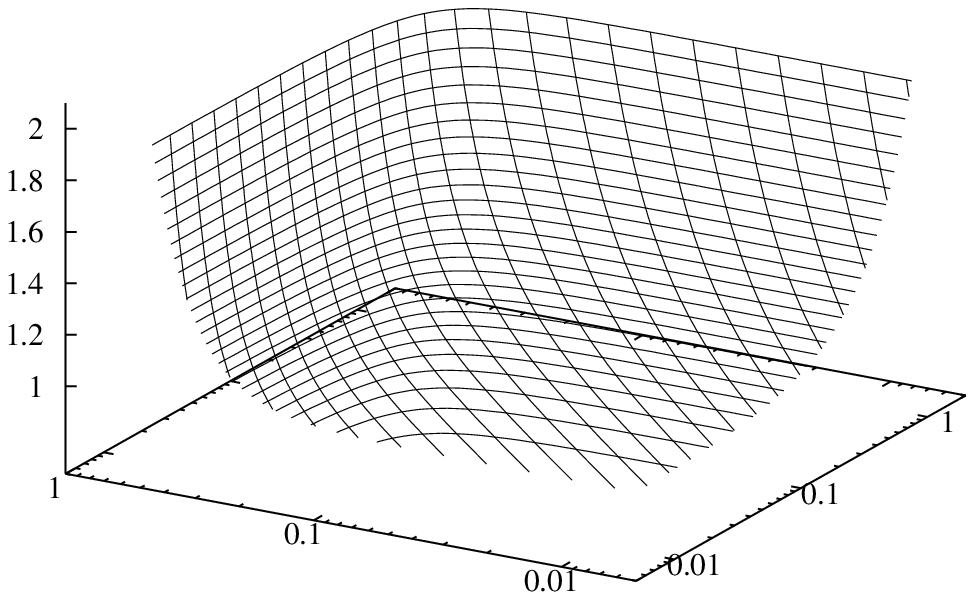}
\put(-290,30){\makebox(0,0){{\Large$\Igi\propto\gd/\sqrt{\sipr}$}}}
\put(-55,50){\makebox(0,0){{\Large$\Ifv\propto\gd/\sipr$}}}
\put(-310,195){\makebox(0,0){{\Large$\sidev/\sipr$}}}
}
\end{center}
\caption{Predicted response for a granular material
in the crossover between the grain inertia regime
and the fluid viscosity regime.
$\sidev/\sipr$ is the effective friction.
$\Igi$ and $\Ifv$ are given
by Eq.~(\ref{Eq:I_gd_sipr_adim}).
The ``gi'' and ``fv'' limits 
($\sidev/\sipr$ as a function of $\Igi$
or as a function of $\Ifv$),
represented on Fig.~\ref{Fig:fonctions},
are (approximately) given by the edges
of the meshed surface.
}
\label{Fig:interpolation}
\end{figure}
}

In the present two-grain approach,
we will use this result with a stress
$\si=\sidev\pm\sipr$ instead of $\sipr$.
Moreover, let us define:
\bee
\label{Eq:Tc}
\Tc &=& \frac{\masse}{\visc\,\rayon} \\
\sic &=& \frac{\visc^2\,\rayon}{\masse}
\label{Eq:sic}
\eee
which are the values taken respectively by $T$ and $\sipr$
at the crossover between both regimes,
see Eqs.~(\ref{Eq:T_gi}) and~(\ref{Eq:T_fv}).

With these notations,
Eq.~(\ref{Eq:solution_equadif})
provides the time at which a grain
submitted to a stress $\si$
has traveled a distance $\rayon$
to meet another grain:
\bee
\label{Eq:si_function_T}
\frac{\sic}{\si} &=& f(T/\Tc) \\
f(x) &=& x-1+e^{-x}
\label{Eq:function_f}
\eee

The lifetime of a contact, 
given by Eq.~(\ref{Eq:shear_rate_related_to_contact_lifetime}),
can be rewritten using Eq.~(\ref{Eq:si_function_T}):
\bee
\frac{1}{\gd\,\Tc}
&=& \frac{\Tapp}{\Tc} + \frac{\Tsep}{\Tc} \nonumber\\
&\simeq& g\left(\frac{\sic/\sipr}{\mu+1}\right)
+ g\left(\frac{\sic/\sipr}{\mu-1}\right)
\label{Eq:shear_rate_related_to_contact_lifetime_interpolated}
\eee
where $\mu=\sidev/\sipr$
and where $g$ is an approximation
(precise up to within two percent)
for the inverse of function $f$
defined by Eq.~(\ref{Eq:function_f}):
\be
\label{Eq:function_g}
g(x)=x+1-e^{-(\sqrt{2x}+x/3)}\simeq f^{-1}(x)
\ee

With the notations
of Eqs.~(\ref{Eq:Tc}-\ref{Eq:sic}),
both limits of the dimensionless parameter $I$
can be expressed as:
\be
\Igi=\frac{\gd\,\Tc}{\sqrt{\sipr/\sic}}
\hs
\label{Eq:I_gd_sipr_adim}
\Ifv=\frac{\gd\,\Tc}{\sipr/\sic}
\ee
Hence, in order to show how it is possible
to interpolate between 
Eqs.~(\ref{Eq:rheo_gi_complete}) and~(\ref{Eq:rheo_fv}),
let us use Eq.~(\ref{Eq:shear_rate_related_to_contact_lifetime_interpolated})
to plot $\mu=\sidev/\sipr$
as a function of both 
$\gd\,\Tc/\sqrt{\sipr/\sic}$
and $\gd\,\Tc/(\sipr/\sic)$.
This is shown on Fig.~\ref{Fig:interpolation}.

\section{Conclusion}

A recent two-grain argument
provided two distinct constitutive laws
for the rheology of granular media in dense regimes:
one for the regime (gi) where the grain inertia is dominant,
and one for the regime (fv) governed by the fluid viscosity.
In the present note, we followed the same arguments
to derive an interpolation between both laws.

Further studies are now needed in order to
{\em (i)} test whether the distinct predictions
are compatible with the experiments,
and {\em (ii)} develop theoretical arguments
that could rationalize the saturation
of the apparent frictional response
as the collisional regime is approached.

\subsection*{Acknowledgements}

This work was supported by the Agence Nationale de la Recherche (ANR05).

\bibliography{fluche,./../../Bibliographie}

\end{document}